# Contact Stiffness and Damping of Liquid Films in Dynamic Atomic Force Microscopy

Rong-Guang Xu and Yongsheng Leng*

*Department of Mechanical and Aerospace Engineering, The George Washington University, Washington, DC 20052, USA*

Small-amplitude dynamic atomic force microscopy (dynamic-AFM) in a simple nonpolar liquid was studied through molecular dynamics simulations. We find that within linear dynamics regime, the contact stiffness and damping of the confined film exhibit the similar solvation force oscillations, and they are generally out-of-phase. For the solidified film with integer monolayer thickness, further compression of the film before layering transition leads to higher stiffness and lower damping. We find that molecular diffusion in the solidified film was nevertheless enhanced due to the mechanical excitation of AFM tip.

---

The mechanical properties of nanoconfined liquid films between two solid surfaces have been long-standing interests in surface force studies. This is largely because of their fundamental importance in surface and interfacial science, and their direct practical applications in friction and lubrication, and adhesion and wear[1-4]. Over the past decades, the squeezing and shear properties of simple liquids [such as octamethylcyclo-tetrasiloxane (OMCTS), a nonpolar model liquid widely used in surface force experimental studies], have been extensively investigated by the surface force apparatus (SFA) or surface force balance (SFB) instrumentations[5-10]. A prevailing perception concerning the layering transition of the liquid film under progressive compression is the liquidlike-to-solidlike phase transition at some $n = 7 \sim 8$ monolayer distance[8,9]. However, the mechanical stiffness and dissipative behavior of the film at integer monolayer thickness or during the layering transition is still not well understood. Since 1990s, the atomic force microscope (AFM) began to be widely used to quantify the surface force and mechanical properties of liquid films at much smaller lateral dimension[11-16]. Molecular stiffness and damping were also studied by AFM[17, 18]. With few exceptions[11, 12], most surface force measurements were carried out by using dynamic atomic force microscopy (dynamic-AFM) due to its high-resolution surface characterization, such as the amplitude-modulation (AM-AFM)[19] and the frequency-modulation (FM-AFM)[20] techniques. In small-amplitude AM-AFM, under acoustic or magnetic driving force excitations[19], the measured amplitude and phase change of an AFM cantilever at different tip-surface distance are used to derive the contact stiffness (conservative force gradient) and damping (viscous dissipation) of a nonpolar liquid film, based on the cantilever dynamics[13, 16, 19]. An important requirement of this procedure is that the cantilever dynamics should be in the linear dynamics regime, in which the tip-surface interaction force during small-amplitude vibration can be represented by the linear terms of force gradient and local damping[16, 19]. A major issue in dynamic AFM is that the interaction between the cantilever itself and the surrounding fluid has a large contribution to the overall damping, resulting in a low quality factor in dynamic AFM[19, 21]. So far it has been shown that as the tip-surface distance decreases in OMCTS, the relationship between the contact stiffness and damping of the confined film can be in-phase[14], in-phase or out-of-phase depending on the normal compression rate[15], or out-of-phase for a large range of driving frequencies[16]. Discrepancies between these findings have not yet been well explained.

In this Letter, we report molecular dynamics (MD) simulations for the dynamic AFM in a simple nonpolar liquid. Simulation results unambiguously show that there are distinct oscillations in both contact stiffness and damping of the confined film versus tip-surface distance. The two quantities are generally out-of-phase, such that the solidlike film at integer monolayer thickness has a higher contact stiffness and a lower damping than the intermediate, liquidlike films during the layer transition. The present study opens a new, theoretical modeling approach that would be useful to explore more sophisticated, yet fundamental questions in dynamic-AFM instrumentations[19, 20], such as probing the hydration shell structure and hydration force around charged surfaces, and understanding the high-resolution dynamic-AFM imaging at biological interfaces[22-24].

We perform MD simulations in a simulation cell containing liquid argon and its vapor phase to mimic the AFM ambient environment. The simulation system is composed of an AFM gold tip and a substrate surface that are completely immersed in argon fluid (Fig. 1). The gold tip is connected to a model spring whose elastic constant, $k_z$, represents the bending force constant of AFM cantilever. The spring constant $k_z$ is set to 40 N/m, a relatively hard spring used in dynamic-AFM[25]. One of the intriguing features in our MD simulation is that there is no physical AFM cantilever involved in simulations. Therefore, the issue of cantilever damping in dynamic-AFM[19, 21] does not exist in MD simulation. The gold tip has a spherical geometry with a radius of 4.5 nm, exposing (111) surfaces at both top and bottom layers. The gold atoms in the top two layers are treated as a rigid entity, which is connected to the model spring to represent cantilever dynamics. The remaining gold atoms in the tip dynamically evolve with time. For the atomic interactions between argon-argon and argon-substrate, the same Lennard-Jones (LJ) potentials used in our previous studies[26, 27] will be applied in this study (i.e., $\varepsilon_{Ar-Ar} = 0.24$ *kcal/mol* and $\sigma_{Ar-Ar} = 0.34$ nm). For gold, we use the embedded atom method (EAM) potential[28] to describe its dynamics. The LJ parameters for gold-argon dispersive interaction are derived from a universal force field[29], where a simple geometric combining rule applies. Hence, we have $\varepsilon_{Au-Ar} = 0.62$ *kcal/mol* and $\sigma_{Au-Ar} = 0.32$ nm.

The dimensions of the simulation box along the *x*-, *y*-, and *z*-directions are 50 nm, 20 nm and 15 nm, respectively. The molecular system includes 6795 gold atoms, 104,092 argon molecules, and 5159 rigid substrate atoms. We apply periodic boundary conditions to three dimensions, keeping the argon liquid phase surrounded by its two stable vapor phases to mimic dynamic-AFM ambient environment. The Nosé-Hoover thermostat is used to control the temperature of the simulation system at 85 K.

Figure 1 shows the equilibrium configuration of the simulation system after 5 ns MD initial equilibration. Dynamic AM-AFM simulations are performed at different tip-substrate distances. This was achieved by slowly relocating the gold tip at different positions away from the surface, through moving the driving support *C* (Fig. 1). We apply a sinusoidal force, $F(t) = F_0 \cos(\omega t)$, directly on the top two layers of the gold tip to mimic the *magnetic* excitation in AM-AFM[13, 19]. The choice of force magnitude $F_0$ is critical to the proper excitation of AFM gold tip. After careful trial-and-error test, we find $F_0$ = 0.85 nN is a good choice to keep small-amplitude oscillation (smaller than one half of a monolayer thickness of argon film) in linear dynamic regime. The driving angular frequency $\omega$ is chosen at $0.93\omega_0$ to increase force measurement sensitivity[14, 16]. Here, $\omega_0$ is the resonant angular frequency in bulk liquid. We have calibrated the resonance curve of the oscillation amplitude versus the excitation frequency far away from the substrate (the lower inset in Fig. 2). By fitting to the Lorentzian expression[19],

$$A(\omega) = \frac{(F_0/k)\omega_0^2}{\sqrt{(\omega_0^2 - \omega^2)^2 + (\omega\omega_0/Q)^2}} \tag{1}$$

we find that $\omega_0$ = 0.1123 $ps^{-1}$ and the Q factor is around 3.32, a typical value of dynamic-AFM in liquids. In vacuum, the resonance angular frequency $\omega_{0\text{-vac}} = (k_z/m)^{1/2}$ = 0.1338 $ps^{-1}$, where $m$ is the net inertia mass of the gold tip ($m$ = 2.22×10$^{-21}$ kg). The reduction in the resonance frequency in liquid is due to the existence of a boundary layer of the fluid surrounding the tip. The effective mass (including the added mass due to the motion of surrounding fluid) is given by $m^* = k_z/\omega_0^2$ = 1.42 $m$.

For the *magnetic* excitation in small-amplitude AM-AFM, the dynamic equation of motion of the gold tip is given by[19]

$$m^* \frac{d^2z}{dt^2} + \gamma_0 \frac{dz}{dt} + k_z z = F_0 \cos(\omega t) + F_{ts}, \tag{2}$$

where $F_{ts}$ is the tip-surface interaction force and $\gamma_0 = m^*\omega_0/Q$ is the damping in bulk liquid. Note that $F_{ts}$ is a function of distance ($z$) and tip oscillation velocity ($dz/dt$) due to the viscous behavior of the confined film. In the linear dynamics regime, $F_{ts}$ can be represented by the linear terms of the interaction force gradient [or the contact stiffness of the confined film, $k_{int} = (dF_{ts}/dz)|_{z_c}$] and the tip-surface local damping ($\gamma_{int}$) around the average position ($z_c$) of the gold tip, viz., $F_{ts}(z_c + z, dz/dt) = F_{ts}(z_c,0) - k_{int} z - \gamma_{int} (dz/dt)$[16]. The simplified *linear* dynamic equation of motion of the gold tip is then written as

$$m^* \frac{d^2z}{dt^2} + (\gamma_0 + \gamma_{int})\frac{dz}{dt} + (k_z + k_{int})z = F_0 \cos(\omega t). \tag{3}$$

The stable solution to equation (3) is given by $z = z_c + A\cos(\omega t + \delta)$, where $A$ is the tip's oscillation amplitude and $\delta$ is the phase difference between the driving force and the tip response. The relevant expressions for these two quantities are[13]

$$A = \frac{F_0}{\sqrt{(k_{int} + k_z - m^*\omega^2)^2 + (\gamma\omega)^2}} \tag{4}$$

$$\tan\delta = \frac{\gamma\omega}{k_{int} + k_z - m^*\omega^2}, \tag{5}$$

in which the total damping $\gamma$ is defined as $\gamma = \gamma_0 + \gamma_{int}$. In MD simulations, we measure $A$ and $\delta$ by fitting $z = z_c + A\cos(\omega t + \delta)$ to the tip oscillation curve. From the above two equations, $k_{int}$ and $\gamma$ are derived as

$$k_{int} = -\frac{F_0}{A}\cos\delta - k_z + m^*\omega^2 \tag{5}$$

$$\gamma = \gamma + \gamma_{int} = -\frac{F_0 \sin\delta}{A\omega} \tag{6}$$

We first investigate the variation of the *averaged* solvation force versus distance in dynamic AM-AFM. This was achieved by averaging the sinusoidal spring force at steady-state excitations of the gold tip at different gap distances. Figure 2 shows the overall solvation force-distance curve, which exhibits an oscillatory feature with a period of molecular diameter of argon (i.e., ~3.4 Å). This indicates that the confined film is squeezed out layer-by-layer between the gold tip and substrate. The upper two insets in the figure show the detailed steady-state force variations with time at $n = 2$ and $n = 6$ monolayer thickness. The lower inset provides the resonance curve (oscillation amplitude versus excitation frequency) of the gold tip far away from the substrate. We find that the oscillatory solvation force cannot be described by a simple oscillatory-decay model (the so-called structural force)[2]. An extra monotonic repulsive force term is needed, as discussed in previous hydration force modeling[22]. As such, the force-distance curve can be fitted by the following force model (the blue line in the figure)

$$F(D) = F_1 \cos\left(\frac{2\pi D}{\sigma}\right)e^{-\frac{D}{\tau_1}} + F_2 e^{-\frac{D}{\tau_2}}, \tag{7}$$

in which the fitted force magnitudes are $F_1 = -48.42$ nN and $F_2 = 122$ nN, and the first and second decay lengths are $\tau_1 = 4.01$ Å and $\tau_2 = 2.9$ Å, respectively. We attribute the second strong repulsive term to the progressively increased density of the solidified film in the last a few layers. To compare with SFA experiments, we recall that the original oscillatory-decay model (the first term in equation (7)) used in SFA was proposed to characterize the force oscillations obtained through continuous *approach-retraction* force measurements[2, 5, 9]. The very slow compression resulted in progressively dense solidified films, while during slow retraction more negative force peaks (adhesion valleys) were explored due to adhesion hysteresis[5, 26]. For such force measurements, the oscillatory-decay model was appropriate. In dynamic-AFM, however, the steady-state vibration of AFM cantilever in fact explores the *average-equilibrium* solvation force at different tip-surface distances. Because of this, the adhesion force will no longer be a part of solvation force. Therefore, the second repulsive term as shown in equation (7) should be considered. In Fig. 2 we also plot the static-AFM force-distance curve during continuous approach (approach speed $v = 1$ m/s). The figure clearly shows that the solvation force curve during continuous compression is higher than the one from dynamic-AFM, especially in the last three solidified layers.

Figure 3 panel A shows the measured amplitude $A$ and phase difference $\delta$ through MD simulations. At larger distances close to the bulk fluid, the amplitude is around 1.7 Å and the initial phase difference between the tip movement and magnetic excitation is around $-60°$. As the tip-surface distance is gradually decreased, both A and $\delta$ begin to decrease and exhibit oscillations with a period of molecular layer thickness of argon. At each integer monolayer distance, as indicated by the number of layers in Fig. 3, amplitude $A$ tends to further decrease as the solidified film becomes more compact under compression, while the magnitude of phase $\delta$ also decreases (symbolizing the solidlike mechanical response). It is only during the layering transition where the solidified film becomes liquidlike, that both $A$ and $\delta$ exhibit dramatic increase, leading to overall oscillations of the two quantities versus tip-surface distance. However, in the last three layers, discontinuous jump occurs for both variables at the point of layering transition. We attribute this to the solvation force gradient exceeding the spring force constant ($k_z = 40$ N/m). Consequently, the gold tip cannot explore these unstable transitions.

Figure 3 panel B shows the derived contact stiffness $k_{int}$ and total damping $\gamma$ of the confined film, based on the measured amplitude $A$ and phase difference $\delta$ (equations (5) and (6)). The figure unambiguously shows that the two variables are out-of-phase, the maxima of the contact stiffness corresponds to the minima of the damping, for the tip-surface distance beyond $n = 3$ layers. At larger distances, $\gamma$ approaches the bulk damping $\gamma_0$. In general, for the solidified film with integer monolayer thickness, further compression of the film before layering transition leads to higher stiffness and lower damping. This viscoelastic behavior is consistent with what has been measured in recent small-amplitude AM-AFM experiment in OMCTS[16], and even in general agrees with the small-amplitude FM-AFM dissipation measurement in $n$-dodecanol film[25].

We emphasize that because of the linear dynamics assumption[19], as indicated in equation (3), we will not be able to use the spring $k_z = 40$ N/m to characterize the contact stiffness and damping for $n = 1$ - 3 layers, to which much stiffer spring is required. In fact, in Fig. 2 we note that the steady-state force variation for $n = 2$ layer already contains noticeable anharmonic component near the shoulder and bottom of the force oscillation, which is in contrast to the well-defined harmonic force oscillation for $n = 6$ layer. Since the solvation force measurement in MD simulation does not take any linear dynamics assumption, as such, its force model (as described by equation (7)), as well as its derivative (the contact stiffness) will represent the realistic properties of the confined film. As shown in the inset in Fig. 3 panel B, in which the contact stiffness $k_{int}$ and total damping $\gamma$ are plotted over the whole range of tip-surface distance, we see that $k_{int}$ obtained from MD simulation deviates significantly from the force gradient for $n = 1$ - 3 layers, while the total damping $\gamma$ completely diverges over the same range of distance.

To further understand the dynamic behavior of molecules in the confined film in dynamic-AFM, we have further calculated the probability density of the film and the diffusion behavior of molecules underneath the gold tip. Figure 4 panels A and B show the density distributions for the $n = 5$ film and for the $n = 5 \to 4$ layering transition, as the gold tip proceeds stable oscillations. Compared to the density distribution under static-AFM at $n = 5$ layers (the dotted line), the density peaks under dynamic-AFM are widened, especially for the top 2-3 layers close to the tip surface, to which the layered structure is more or less disrupted due to the perturbation of tip movements. In Fig. 4 panel C we compare the mean square displacements (MSDs) of argon molecules under static- and dynamic-AFM, for different confined films. For the very dense, solidified film at $n = 3$ monolayer thickness and below, the oscillation of gold tip does not change the diffusion behavior of confined molecules. For the solidified film at $n = 4 - 5$ monolayer thickness, the effect of tip oscillation gradually enhances the diffusion of confined molecules, indicating that they are more or less excited by mechanical perturbations. MSD curves for $n = 6 - 8$ monolayers under static-AFM are also shown in panel C, from which we anticipate that these MSDs under dynamic-AFM will shift dramatically towards the bulk MSD (not shown in the panel).

While the correlation between the contact stiffness and damping of simple nonpolar liquid under confinement is still under debate[14-16], the present work suggests that, within the linear dynamics regime, they are generally out-of-phase. The origin of this correlation has been attributed to the fluctuations of the number of molecular layers: the elastic property (at integer layers) and viscous behavior (during $n \to n$-1 transition) of the confined film[30]. A physical picture to explain this is related to the two different regimes for the dissipation of vibration energy of an oscillating AFM tip[25]. As shown in the inset in Fig. 4 panel A, when an integer-monolayer film ($n = 5$) undergoes cyclic elastic compression without disruption of the layered structure, minimal reconfiguration of molecules is needed, thus giving high stiffness with low damping. On the other hand, upon the layering transition (the inset in Fig. 4 panel B), repeated squeeze out and reformation of a monolayer will result in large energy dissipation and damping.

This work was supported by the National Science Foundation (NSF 1149704) and the National Energy Research Scientific Computing Center (NERSC).

[*] Email: leng@gwu.edu

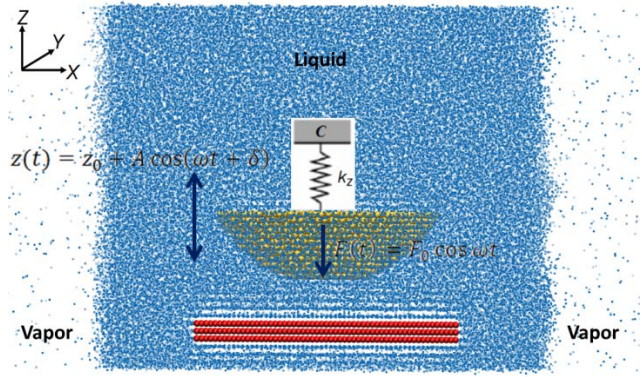

**FIG. 1** (color online). Schematic of the MD simulation system. The AFM gold tip is connected to a driving support $C$ by a spring ($k_z$). Both tip and substrate surface are immersed in argon liquid, which is surrounded by a stable argon vapor phase to mimic dynamic-AFM force measurement. Under magnetic excitation, $F(t) = F_0 \cos(\omega t)$, the gold tip proceeds a simple harmonic oscillation in liquid.

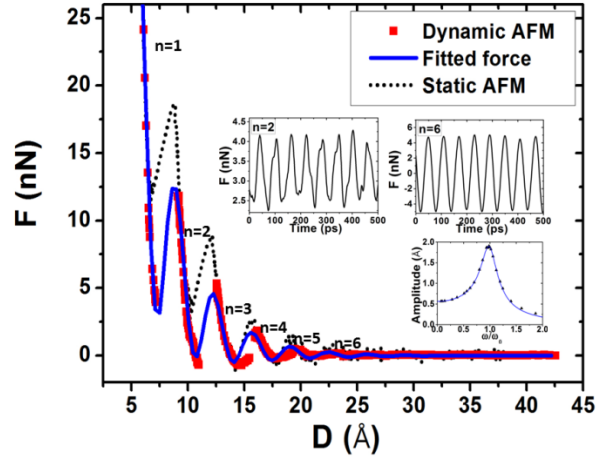

**FIG. 2** (color online). The averaged solvation force versus distance in dynamic AM-AFM. The static AFM force curve is also shown in the figure for comparison. The lower inset shows the Lorentzian resonance curve of AFM tip in the bulk liquid. The upper two insets show the steady-state force variations at $n = 2$ and $n = 6$ monolayer thickness.

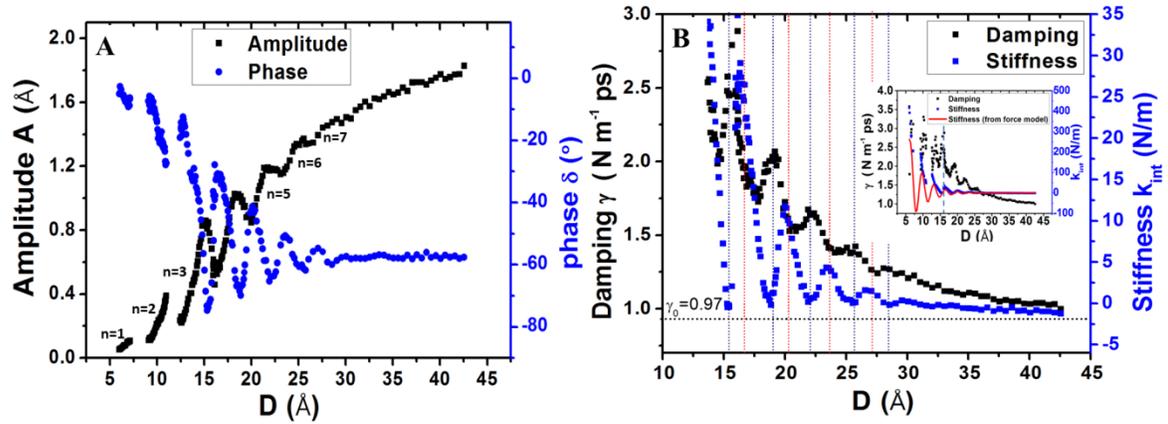

**FIG. 3** (color online). **A.** Variations of the amplitude $A$ and phase difference $\delta$ from MD simulations. **B.** the derived contact stiffness $k_{int}$ and total damping $\gamma$ of the confined film, based on the measured $A$ and $\delta$. The inset in panel B shows that both $k_{int}$ and $\gamma$ diverge for $n = 1 - 3$ layers, due to the nonlinear dynamics behavior of the gold tip.

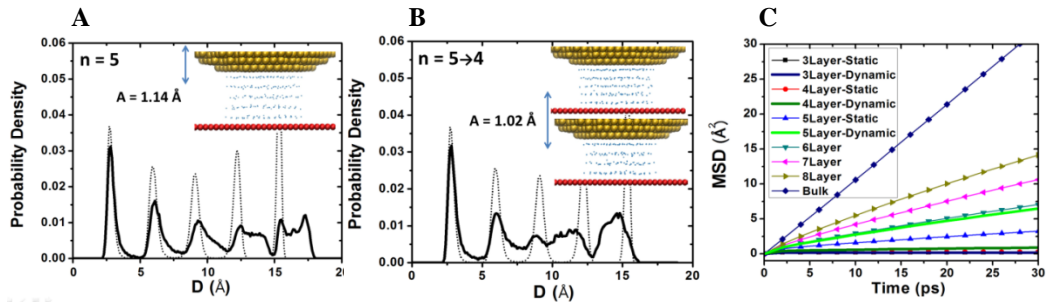

**FIG. 4** (color online). The probability density distributions for the $n = 5$ film **(A)** and for the $n = 5 \rightarrow 4$ layering transition **(B)**, as the gold tip proceeds stable oscillations. Density distributions under static-AFM at $n = 5$ layers (the dotted line) are also shown for comparison. The insets in the two panels show the detailed molecular configurations during the tip oscillation. **(C)** The mean square displacements (MSDs) of argon molecules under static- and dynamic-AFM. The MSD curve for the bulk fluid is also shown in the figure for comparison.